\documentclass[a4paper,aps,preprint,prl,nofootinbib,twocolumn,reprint]{revtex4-1}
\usepackage{amsmath,amsthm}
\usepackage{comment,amssymb,latexsym,upref}

\newcommand{\mnote}[1]{}

\newcommand{\eqn}[2]{ \begin{equation*} #2  \end{equation*} }

\newcommand{\lalign}[2]{ \begin{align} #2 \end{align} }
\newcommand{\alin}[2]{ \begin{align*} #2 \end{align*} }
\newcommand{\leqn}[2]{\begin{equation} #2 \label{#1} \end{equation}}

\newcommand{\sect}[2]{\section{#2}\label{#1}}

\theoremstyle{plain}

\theoremstyle{definition}

\theoremstyle{remark}

\DeclareMathAlphabet{\mathpzc}{OT1}{pzc}{m}{it}

\newcommand{\id}{\mathord{{\mathrm 1}\kern-0.27em{\mathrm I}}\kern0.35em}
\newcommand{\del}{\partial}
\newcommand{\delb}{\bar{\partial}}

\newcommand{\Half}{\ensuremath{\textstyle\frac{1}{2}}}
\newcommand{\Quarter}{\ensuremath{\textstyle\frac{1}{4}}}

\newcommand{\Twothirds}{\ensuremath{\textstyle\frac{2}{3}}}

\newcommand{\norm}[1]{\|#1\|}

\newcommand{\nt}{\tilde{n}{}}

\newcommand{\alphat}{\tilde{\alpha}}
\newcommand{\pt}{\tilde{p}}

\newcommand{\rhot}{\tilde{\rho}}
\newcommand{\mut}{\tilde{\mu}}

\newcommand{\Phit}{\tilde{\Phi}}

\newcommand{\Psit}{\tilde{\Psi}{}}

\newcommand{\gb}{\bar{g}{}}

\newcommand{\Gb}{\bar{G}{}}

\newcommand{\pb}{\bar{p}{}}

\newcommand{\Tb}{\bar{T}{}}
\newcommand{\vb}{\bar{v}{}}

\newcommand{\xb}{\bar{x}{}}

\newcommand{\rhob}{\bar{\rho}{}}

\newcommand{\nablab}{\bar{\nabla}{}}

\newcommand{\gh}{\hat{g}{}}
\newcommand{\nh}{\hat{n}{}}

\newcommand{\alphah}{\hat{\alpha}{}}

\newcommand{\muh}{\hat{\mu}{}}
\newcommand{\Phih}{\hat{\Phi}}

\newcommand{\Psih}{\hat{\Psi}}

\newcommand{\Rf}{\mathfrak{R}}

\newcommand{\xv}{\mathbf{x}}
\newcommand{\yv}{\mathbf{y}}

\newcommand{\Rbb}{\mathbb{R}}

\newcommand{\Tbb}{\mathbb{T}}

\newcommand{\ep}{\epsilon}

\begin{document}

\title{Cosmological Newtonian limit}
\author{Todd A. Oliynyk}
\affiliation{School of Mathematical Sciences\\
Monash University, Victoria 3800\\
Australia}
\email{todd.oliynyk@monash.edu}

\begin{abstract}
We describe a wide class of inhomogeneous relativistic solutions that
are well approximated on cosmological scales by solutions of Newtonian gravity. Error estimates
measuring the difference between the Newtonian and relativistic solutions are given.
The solutions presented here unambiguously demonstrate that it is possible for Newtonian gravity to provide
a good approximation to General Relativity on cosmological scales.
\end{abstract}

\maketitle

\sect{intro}{Introduction}
In recent years, there has been a resurgence in interest in the relationship between Newtonian gravity and
General Relativity on cosmological scales
\cite{BuchertRasanen:2012,Ellis:2011,Clarkson_etal:2011,Green_Wald:2011,Green_Wald:2012,Hwangetal:2008,HwangNoh:2013,MatarreseTerranova:1996,Rasanen:2010},
which is motivated by
questions surrounding the physical interpretation of large scale cosmological simulations  using Newtonian gravity and
the role of Newtonian gravity in cosmological averaging. The overriding fundamental question is
in what sense, if any, does Newtonian gravity provide an approximation to General Relativity on a cosmological
scale.

At the level of field equations, it is well known that
the relationship between Newtonian gravity and General Relativity can be established through the introduction of a small parameter $\ep=v/c$, where
$v$ is a typical speed of the gravitating matter and $c$ is the speed of light, into the Einstein-matter field
equations. By assuming an appropriate dependence of the metric and matter fields on $\ep$, the Newtonian-matter
field equations are then recovered in the \emph{singular} limit $\ep \searrow 0$. Although this type of formal calculation is
transparent, it does not provide any real answers because the physics is governed by solutions of the Einstein-matter
equations, and consequently, it is the solutions that must be examined in the limit $\ep \searrow 0$.
In general, this is a much more difficult task;
in the articles \cite{Oliynyk:CMP_2010,Oliynyk:JHDE_2010}, we introduced an approach to rigorously analyze this problem with the goal
of constructing 1-parameter families of $\ep$ dependent solutions to the Einstein-Euler equations
that limit, as $\ep \searrow 0$, to solutions of the cosmological Poisson-Euler equations of Newtonian
gravity. Although we were able to construct such solutions, it was pointed out in \cite{Green_Wald:2012}
that the class of solutions we produced are not valid on a cosmological scale. As a consequence, the results of the
article \cite{Oliynyk:CMP_2010,Oliynyk:JHDE_2010} do not directly address the relationship between
solutions of Newtonian and Einstein gravity on a cosmological scale; however, the fact that
a large class of solutions of the cosmological Poisson-Euler equations were shown to be limits
of relativistic solutions  strongly suggests that Newtonian gravity can provide a good approximation
to fully relativistic solutions on a cosmological scale.

In this article, we describe improvements to the approach used in \cite{Oliynyk:CMP_2010,Oliynyk:JHDE_2010}
that address the deficiencies pointed out in \cite{Green_Wald:2012}. The result being
that we can prove the existence of 1-parameter families of solutions to the Einstein-Euler equations that are uniformly approximated
on \emph{cosmological scales}
by solutions of the cosmological Poisson-Euler equations. We will not dwell here on
the technical aspects of the construction, and instead concentrate on describing the solutions,
and more importantly, the way in which they approximate relativistic cosmological solutions. Technical details for
the construction will be provided in a forthcoming article \cite{Oliynyk:2013}.

In order to describe our new 1-parameter families of solutions and the improvement over the previous results \cite{Oliynyk:CMP_2010,Oliynyk:JHDE_2010}, we first fix our notation. We let $\mathbf{\xb}=(\xb^I)$ $(I=1,2,3)$ denote global Cartesian coordinates
on $\Rbb^3$, which can be thought of as ``spatial'' coordinates. We also assume that
spacetime is a slab of the form $M=[0,T]\times \Rbb^3$ on which $\xb= (\xb^i)=(\xb^0,\xb^I)$
are global Cartesian coordinates. We refer to the $(\xb^i)$ as \emph{relativistic coordinates}. These coordinates are
characterized by the requirement that the spacetime metric $\gb=\gb_{ij}d\xb^i d\xb^j$ converges, in these coordinates, to a well-defined Lorentzian metric in
the limit $\ep \searrow 0$.

By suitably rescaling the spacetime metric $\gb=\gb_{ij}d\xb^i d\xb^j$ and fluid variables $\{\rhob,\vb=\vb^i\delb_i\}$, the
(isentropic) Einstein-Euler equations, in the coordinates  $(\xb^i)$, can be written in the following non-dimensionalized form
\lalign{Eineul}{
\Gb^{ij} &= 2\bigl( \Tb^{ij}-\Lambda \gb^{ij}\bigr), \label{Eineul.1} \\
\nablab_i \Tb^{ij} & = 0 \label{Eineul.2},
}
where
\eqn{stress}{
\Tb^{ij} = (\rhob + \pb)\vb^i\vb^j + \pb \gb^{ij}
}
is the stress energy tensor of an isentropic perfect fluid with equation
of state
\eqn{eos}{
\pb=\ep^2 f(\rhob),
}
and the fluid four-velocity $\vb^i$ is normalized according
to
\eqn{vnorm}{
\vb^i\vb_i = -1.
}

In the more familiar and well studied setting of isolated systems, we previously \cite{Oliynyk:CMP_2007} established the existence of 1-parameter families of solutions
$\{\gb^\ep_{ij},\vb^i_\ep,\rhob_\ep\}$, $0<\ep\leq\ep_0$, to the Einstein-Euler equations (\ref{Eineul.1}-\ref{Eineul.2}) on $M=[0,T]\times \Rbb^3$ that
converge, in a suitable sense, to solutions of the Poisson-Euler equations of Newtonian gravity. In the relativistic coordinates, the metric
$\gb^\ep_{ij}$ for these solutions converges uniformly to the Minkowski metric $\eta_{ij}$ as $\ep \searrow 0$, while the density $\rhob_\ep$ is of characteristic
size $\sim \ep$ and concentrates, when suitably rescaled, to a Dirac delta function about its center of mass in the limit $\ep \searrow 0$. This agrees
with the expected behavior for the Newtonian limit of isolated systems. For readers unfamiliar with this viewpoint
on the Newtonian limit, a more traditional viewpoint is achieved by transforming to \emph{Newtonian coordinates} $(t,x^I)$, which are related to the relativistic coordinates
through the simple scaling relation
\eqn{coords}{
\xb^0 = t, \quad \xb^J = \ep x^J.
}
Expressed in these coordinates, the metric $g^\ep_{ij}$ converges to a degenerate two tensor, while the density $\rho_\ep$ and spatial components of the fluid velocity $v^I_\ep$ converge
uniformly to solutions of the Poisson-Euler equations.

In the articles \cite{Oliynyk:CMP_2010,Oliynyk:JHDE_2010}, we established the existence of 1-parameter families of solutions $\{\gb^\ep_{ij},\vb^i_\ep,\rhob_\ep\}$, $0<\ep\leq\ep_0$ to the Einstein-Euler
equations (\ref{Eineul.1}-\ref{Eineul.2}) on $M_\ep \cong [0,T]\times \Tbb^n$ that converge to solutions of the cosmological Poisson-Euler equations
as $\ep \searrow 0$. Lifting these to the covering
space, they become periodic solutions of (\ref{Eineul.1}-\ref{Eineul.2}) on $M \cong [0,T]\times \Rbb^n$ with period $\sim \ep$.
This leads, as discussed in \cite{Green_Wald:2012}, to the interpretation of these solutions as being local since the ``size'' of the torus is $\sim \ep$, and hence, shrinks to zero in the limit $\ep \searrow 0$
in complete analogy with the behavior of isolated systems where the matter is of characteristic size $\sim \ep$ and shrinks to
zero in the limit $\ep \searrow 0$. One concludes from this that the solutions from \cite{Oliynyk:CMP_2010,Oliynyk:JHDE_2010} do not represent fully relativistic solutions
that converge on cosmological scales to solutions of the cosmological Poisson-Euler equations, but instead, represent solutions that converge on scales comparable to isolated systems.

Our approach to extend the results from \cite{Oliynyk:CMP_2010,Oliynyk:JHDE_2010} to cosmological scales is to essentially ``glue'' together multiple, spatially separated solutions that have local behavior
similar to the solutions from \cite{Oliynyk:CMP_2010,Oliynyk:JHDE_2010}. The main difficulty is keeping the local solutions separated in the limit $\ep \searrow 0$ by a finite
light travel time, which is absolutely necessary for the total solution to be cosmologically relevant. Keeping this separation is difficult because it is in conflict with basic assumption of the Newtonian limit, which is that the speed of light, relative to a characteristic matter speed,
goes to infinity. In this sense, the Newtonian limit in the cosmological setting is even more singular than the standard Newtonian limit.

The key to achieving the required separation is in the correct choice of initial data. As we
describe below, the cosmological Newtonian limit requires that the initial data for \emph{both} the Einstein-Euler equations and the cosmological Poisson-Euler equations be chosen
in an $\ep$-dependent fashion. This is in stark contrast to the standard Newtonian limit where the initial data for the Poisson-Euler equations is $\ep$-independent. One immediate consequence
is that, in the cosmological setting, there is no global convergence as $\ep \searrow 0$ for the matter fields in the Newtonian coordinates. Instead, there is
local convergence around the center of mass of each of the local systems to a solution of the cosmological Poisson-Euler equations.

\sect{cosm}{Cosmological Newtonian solutions}

The starting point for our construction is to produce cosmologically relevant solutions
of the cosmological Poisson-Euler equations given by
\lalign{Eulexp0aa}{
\del_0 \mut + \nt^I\del_I\mut  +  \rhot \del_I \nt^I &= 3\mut\Psit^{\prime}, \label{Eulexp0aa.1}\\
\rhot\bigl(\del_0 \nt^J+\nt^I\del_I \nt^J\bigr) +
f'\bigl(\rhot\bigr)\del^J\mut
 &= - \Quarter \rhot\Phit^J +\rhot\nt^J\Psit^\prime ,\label{Eulexp0aa.2} \\
 \del_0 \bigl(e^{-\Psit}\Phit_I\bigr) + 4 \Rf_I\Rf_J \Bigl(e^{\Psit}\rhot\nt^J\Bigr)&=0,
\label{Eulexp0aa.3} \\
\alphat^\prime - 3\Psit^{\prime}\alphat &= 0, \label{Eulexp0aa.4} \\
\Psit^{\prime\prime}-\Half\bigl(\Psit^{\prime}\bigr)^2 &= -e^{-2\Psit}\Lambda ,\label{Eulexp0aa.5}
}
where $(\cdot)^\prime = d(\cdot)/dt$ denotes differentiation with respect to time, $(t,\xv)=(t,x^I)$ are
Newtonian coordinates, $\del_I=\del/\del x^I$, $\Delta=\delta^{IJ}\del_I\del_J$ is the Euclidean
Laplacian, $\Rf_I = -(-\Delta)^{-1/2}\del_I$
is the Riesz transform\footnote{The operator $(-\Delta)^{-\frac{1}{2}}$ can be computed via the
formula $ (-\Delta)^{-\frac{1}{2}}(f)(\xv) =
\frac{1}{2\pi^2}\int_{\Rbb^3} \frac{f(\yv)}{|\xv-\yv|^{2}} d^3 \yv$
for functions $f(\xv)$ that decay sufficiently fast near infinity.}, and the spatial indices (i.e. $I,J,K$) are raised and lowered with the Euclidean metric $\delta_{IJ}$.
In this formulation
\eqn{Eulexp0ab}{
\rhot(t,\mathbf{x}) = \alphat(t) + \mut(t,\mathbf{x})
}
is the mass density, the pressure $\pt$ is determined by the equation of state
\eqn{neos}{\pt=f(\rhot),
}
$\Lambda$ is the cosmological constant, $\Psit(t)$ is a time dependent conformal factor that accounts for the
expansion of space, and $n^I$ is the conformally rescaled fluid 3-velocity.

The initial data for (\ref{Eulexp0aa.1}-\ref{Eulexp0aa.5}) separates into
free and constrained components. For the free initial data, we choose
\lalign{idmatAa}{
&\alphat(0) = \alphah, \label{idmatAa.1} \\
&\Psit(0) = \Psih, \label{idmatAa.2}\\
&\mut(0,\mathbf{x}) = \muh_{y,\ep}(\mathbf{x}) := \sum_{a=1}^N \muh_a(\mathbf{x}-\mathbf{y}_a/\ep),  \label{idmatAa.3}\\
&\nt^I(0,\mathbf{x}) = \nh^I_{y,\ep}(\mathbf{x}) :=
 \frac{1}{\alphah+\muh_{y,\ep}(\mathbf{x})} \times \notag\\
&\Biggl(
 \sqrt{\Twothirds e^{-2\Psih}(\alphah+\Lambda)}
  \del^{I}\Delta^{-1}\muh_{y,\ep}(\mathbf{x})+\sum_{a=1}^N\nh_a^I(\mathbf{x}-\mathbf{y}_a/\ep)\Biggr),
  \label{idmatAa.4}
}
where $\Delta^{-1}$ is the inverse Laplacian\footnote{Given explicitly by
$\Delta^{-1}(f)(\xv) = -\frac{1}{4\pi} \int_{\Rbb^3} \frac{f(\yv)}{|\xv-\yv|} d^3 \yv$ for functions $f(\xv)$ that decay
sufficiently fast near infinity.}, $\alphah > 0$, $\Psih\in \Rbb$, $\ep > 0$, $y= (\mathbf{y}_1,\ldots,\mathbf{y}_N) \in (\Rbb^{3})^N$,
$\muh_a, \nh^I_a \in L^{6/5}(\Rbb^3)\cap H^{s+1}(\Rbb^3)$ with $s>3/2+1$,
and the $\alphah$ and $\muh_a$  are chosen so that
\eqn{idmatC}{
\rhot(0,\mathbf{x})= \alphah + \mu_{y,\ep}(\mathbf{x}) \geq c_0 > 0 \quad \forall\; \mathbf{x}\in \Rbb^3,
\; \ep > 0,\; y\in \Rbb^{3N}
}
for some constant $c_0$. The constrained initial data is determined in terms of the free data by
\leqn{idmatAb.1}{
\Psit^{\prime}(0) = \sqrt{\Twothirds e^{-2\Psih}(\alphah+\Lambda)}
}
and
\leqn{idmatAb.2}{
\Phit_I(0,\mathbf{x}) = \del_I \Phih(\mathbf{x}),
}
where $\Phih$ is a solution
\eqn{idmatAc}{
\Delta \Phih(\mathbf{x}) = 4e^{-2\Psih}\mut(0,\mathbf{x}).
}
We note that the terminology of constrained initial data is appropriate as it can be shown that the initial
data constraints (\ref{idmatAb.1}-\ref{idmatAb.2}) are preserved under evolution; that is, solutions
of (\ref{Eulexp0aa.1}-\ref{Eulexp0aa.5}) subject to (\ref{idmatAb.1}-\ref{idmatAb.2}) satisfy
\leqn{idmatAc}{
\Psit^{\prime} = \sqrt{\Twothirds e^{-2\Psit}(\alphat+\Lambda)}
}
and
\eqn{idmatAca}{
\Phit_I = \del_I \Phit,
}
where $\Phit$ solves
\eqn{idmatAcb}{
\Delta \Phit = 4e^{-2\Psit}\mut.
}

We also observe that the evolution equations (\ref{Eulexp0aa.4}-\ref{Eulexp0aa.5}) subject to the initial
data constraint \eqref{idmatAb.1} are completely equivalent to the equations of motion satisfied by a  Friedmann-Lema\^{i}tre-Robertson-Walker (FLRW) dust solution
with a
cosmological constant $\Lambda$. To see that this is the case, we define a conformal
related time parameter $\tau=\tau(t)$ by
\eqn{flrwA}{
\tau = \int_0^t e^{-\Psit(s)}\,ds,
}
and set $a(\tau) = e^{-\Psit(t(\tau))}$ and $\alpha(\tau)=\alphat(t(\tau))$. A straightforward calculation
using (\ref{Eulexp0aa.4}-\ref{Eulexp0aa.5}) and \eqref{idmatAc} then shows that $\{a(\tau),\alpha(\tau)\}$ satisfy
\alin{flrwB}{
3\left(\frac{\dot{a}}{a}\right)^2 &= 2(\alpha + \Lambda) \qquad \bigl( \dot{a}=da/d\tau \bigr), \\
3\frac{\ddot{a}}{a} &= -(\alpha - 2\Lambda),
}
which agrees with the standard presentation of the FLRW dust equations with a cosmological constant.

At the initial time $t=0$ and for fixed $\ep >0$, the initial data
(\ref{idmatAa.1}-\ref{idmatAb.2})
represents a fluid that initially consists of a homogenous component of size $\alphah$ that is superimposed with $N$ density\footnote{We can even take $N=\infty$
provided that the $\muh_a$ and $\nh_a$ satisfy
$\sum_{a=0}^\infty\bigl( \norm{\muh_a}_{L^{6/5}}+\norm{\muh_a}_{H^{s+1}}\bigr) < \infty$
and $\sum_{a=0}^\infty\bigl( \norm{\nh_a}_{L^{6/5}}+\norm{\nh_a}_{H^{s+1}}\bigr) < \infty$, respectively.}
``fluctuations'' centered at the spatial points $\mathbf{y}_a/\ep \in \Rbb^3$ $(a=1,2,\ldots, N)$ with profiles determined by the functions
$\muh_a$. Moreover, it can be shown that the initial data can be estimated by
\alin{idmatD}{
\norm{\muh_{y,\ep}}_{L^{6/5}} + \norm{\muh_{y,\ep}}_{H^{s+1}}& \leq \sum_{a=1}^N \Bigl(\norm{\muh_a}_{L^{6/5}}
+ \norm{\muh_a}_{H^{s+1}}\Bigr) 
\intertext{and}
\norm{\nh_{y,\ep}}_{L^{6/5}} + \norm{\nh_{y,\ep}}_{H^{s+1}}&
\leq C\bigl(\norm{\muh_{y,\ep}}_{H^{s+1}}\bigr) \sum_{a=1}^N \Bigl(\norm{\muh_a}_{L^{6/5}} \notag \\
&\hspace{-0.7cm}+ \norm{\muh_a}_{H^{s+1}}+ \norm{\nh_a}_{L^{6/5}}
+ \norm{\nh_a}_{H^{s+1}}\Bigr). 
}
A direct consequence of this is that the size of the initial data in the space $L^{6/5}(\Rbb^3)\cap H^{s+1}(\Rbb^3)$ is
independent of the choice of $y\in (\Rbb^{3})^N$ and $\ep > 0$.
Standard existence theorems for hyperbolic equations  then guarantee the existence of a $T>0$, independent of  $y\in (\Rbb^{3})^N$ and $\ep > 0$, and maps
\lalign{cnsol}{
\mut_{y,\ep}, \nt^I_{y,\ep}, \Phit^I_{y,\ep} &\in \bigcap_{\ell=0}^1 C^\ell([0,T],H^{s+1-\ell}(\Rbb^3)),
\label{cnsol.1}\\
\alphat, \Psit &\in C^\infty([0,T],\Rbb), \label{cnsol.2}
}
such that $\{\mut_{y,\ep}, \nt^I_{y,\ep}, \Phit^I_{y,\ep},\alphat, \Psit\}$ solves (\ref{Eulexp0aa.1}-\ref{Eulexp0aa.5}) on the spacetime region $M=[0,T]\times \Rbb^3$ and
agrees at $t=0$ with the initial data (\ref{idmatAa.1}-\ref{idmatAb.2}).


\sect{rel}{Cosmological approximations}

The interpretation of the Newtonian cosmological solutions from the previous section rely on the existence
of a nearby relativistic solution. That such relativistic solutions exists is a consequence of the results of
\cite{Oliynyk:2013}, which are, in turn, based on the ideas developed in the articles
\cite{Oliynyk:CMP_2007,Oliynyk:CMP_2009,Oliynyk:CMP_2010,Oliynyk:JHDE_2010,Oliynyk:ATMP_2012}  where related problems were addressed. For the purposes of this article, the details of the existence proof are not important,
the only relevant fact is that they exist and  are approximated by, in a suitable sense, the Newtonian cosmological solutions from the previous
section.

The basic existence and approximation results from \cite{Oliynyk:2013} may be summarized as follows:
letting $T$ be
as in (\ref{cnsol.1}-\ref{cnsol.2}) and $s>3/2+1$, there exists
an $\ep_0 >0$ such that for every $(y,\ep) \in (\Rbb^3)^N\times (0,\ep_0]$
there exists maps\footnote{Here, $K^{s}(\Rbb)$ $(s\geq 1)$ is the completion of $C_0^\infty(\Rbb^3)$ in the norm
$\norm{u}_{K^s}=\norm{u}_{L^\infty} + \norm{D u}_{H^{s-1}}$.}
\lalign{crelsol}{
&\psi^{y,\ep},\psi^{y,\ep}_0, \mu_{y,\ep}, u^{ij}_{y,\ep} \in \bigcap_{\ell=0}^1([0,T],K^{s-\ell}(\Rbb^3)\cap L^6(\Rbb^3)), \label{cresol.1} \\
&\psi^{y,\ep}_I, n_I^{y,\ep}, \Phi^I_{y,\ep}, u_{y,\ep,k}^{ij} \in \bigcap_{\ell=0}^1 C^\ell([0,T],H^{s-\ell}(\Rbb^3)) ,
\label{cresol.2} \\
&\alpha_\ep, \Psi_\ep \in C^\infty([0,T],\Rbb), \label{cresol.3}
}
that satisfy the estimates
\lalign{crelest}{
&\norm{u^{ij}_{y,\ep}}_{L^\infty([0,T],K^{s-1}\cap L^6)}+
\norm{u^{ij}_{y,\ep,0}}_{L^\infty([0,T],K^{s-1}\cap L^6)} \notag \\
&\hspace{0.6cm} + \norm{u^{ij}_{y,\ep,I}-\delta^i_0\delta^j_0\Phit_{y,\ep,I}}_{L^\infty([0,T],K^{s-1}\cap L^6)}  \leq C\ep, \label{crelest.1}\\
&\norm{\psi^{y,\ep}}_{L^\infty([0,T],K^{s-1}\cap L^6)} \notag \\
&\hspace{2.8cm} +\norm{\psi^{y,\ep}_i}_{L^\infty([0,T],K^{s-1}\cap L^6)}
\leq C\ep, \label{crelest.2}\\
&\norm{\mu_{y,\ep}-\mut_{y,\ep}}_{L^\infty([0,T],K^{s-1}\cap L^6)} +\norm{n_0^{y,\ep}}_{L^\infty([0,T],K^{s-1}\cap L^6)}\notag \\
&\hspace{1.7cm}+ \norm{n_I^{y,\ep}-\nt^I_{y,\ep}}_{L^\infty([0,T],K^{s-1}\cap L^6)}    \leq C \ep, \label{crelest.3}
\intertext{and}
&\norm{\alpha_\ep-\alphat}_{L^\infty([0,T])}+
 \norm{\Psi_\ep-\Psit}_{L^\infty([0,T])} \notag \\
 &\hspace{3.4cm}+\norm{\Psi^\prime_\ep-\Psit^\prime}_{L^\infty([0,T])}  \leq C\ep,
\label{crelest.4}
}
for some constant $C$ independent of $y \in (\Rbb^3)^N$ and $\ep \in (0,\ep_0]$. Moreover,
these maps determine a solution
\alin{crelsolA}{
\gb = \gb^{ij}_{y,\ep}(\xb)\delb_i\delb_j, \quad \vb = \vb_i^{y,\ep}(\xb)d\xb^i, \quad \rhob = \rhob_{y,\ep}(\xb),
}
in relativistic coordinates
to the Einstein-Euler equations (\ref{Eineul.1}-\ref{Eineul.2}) on the spacetime region $M=[0,T]\times \Rbb^3$
according to the formulas:


\lalign{crelsolC}{
\gb^{ij}_{y,\ep}(\xb^0,\mathbf{\xb}) &= \frac{e^{2(\Psi_\ep(\xb^0)+\ep\bar{\psi}^{y,\ep}(\xb^0,\mathbf{\xb}))}}{\sqrt{-\det\bigl(\gh_{y,\ep}^{kl}(\xb^0,\mathbf{\xb})}\bigr)}
\gh^{ij}_{y,\ep}(\xb^0,\mathbf{\xb}),
\label{crelsolC.1} \\
\rhob_{y,\ep}(\xb^0,\mathbf{\xb}) &= \alpha_\ep(\xb^0) + \mu_{y,\ep}(\xb^0,\mathbf{\xb}/\ep), \label{crelsolC.2} \\
\vb_i^{y,\ep}(\xb^0,\mathbf{\xb}) &= e^{-\Psi_\ep(\xb^0)-\ep\bar{\psi}^{y,\ep}(\xb^0,\mathbf{\xb})}
\bigl( -\delta_i^0 + \ep n_i^{y,\ep}(\xb^0,\mathbf{\xb}/\ep)\bigr), \label{crelsolC.3}
}
where
\alin{ghat}{
\gh^{ij}_{y,\ep}(\xb^0,\mathbf{\xb}) &= \eta^{ij} + \ep u^{ij}_{y,\ep}(\xb^0,\mathbf{\xb}/\ep),\\
\delb_k \gh^{ij}_{y,\ep}(\xb^0,\mathbf{\xb}) &= \ep u^{ij}_{y,\ep,k}(\xb^0,\mathbf{\xb}/\ep),\\
\bar{\psi}_{y,\ep}(\xb^0,\mathbf{\xb}) &= \psi^{y,\ep}(\xb^0,\mathbf{\xb}/\ep), \\
\delb_k \bar{\psi}^{y,\ep}(\xb^0,\mathbf{\xb}) &= \psi^{y,\ep}_k(\xb^0,\mathbf{\xb}/\ep),
}
and the conformal harmonic gauge condition $\delb_i\gh^{ij}_{y,\ep} = 0$ is satisfied.

From the estimates (\ref{crelest.1}-\ref{crelest.4}) and the formulas (\ref{crelsolC.1}-\ref{crelsolC.3}), we draw
the following conclusions:
\begin{itemize}
\item[(i)] the Newtonian solutions (\ref{cnsol.1}-\ref{cnsol.2}) uniformly approximate fully relativistic solutions
on the space time region $M=[0,T)\times \Rbb^3$,
\item[(ii)] the metric of the fully relativistic solutions can be viewed as a ``perturbation'' of a FLRW dust solution, and
\item[(iii)] in the \textit{Newtonian coordinates} $(t,\mathbf{x})$ and for $\ep_1 \in (0,\ep_0]$ small enough, the intersection of the backwards light cone emanating from the point $(t,\mathbf{x})=(T,0)$ with the hypersurface $t=0$ contains the ball $B_{T/(2\ep)}(0)$ for all $\ep \in (0,\ep_1]$.
\end{itemize}
A direct consequence of (iii) is that if we set
\eqn{Icdef}{
\mathcal{I} = \{ a\in \{1,2,\ldots,N\}\, |\, |\mathbf{y}_a|\leq T/2 \,\},
}
and fix $\ep \in (0,\ep_1]$, then the point $(T,0)$, in Newtonian coordinates,
contains in its domain of dependence part or all, depending on the size of the support of the $\muh_a$, of the
initial density inhomogeneities $\muh_a(\mathbf{x}-\mathbf{y}_a/\ep)$ centered at $\mathbf{y}_a/\ep$, $a\in \mathcal{I}$. Furthermore,
we note that since $T$ is independent of the choice of $y=(\mathbf{y}_1,\ldots,\mathbf{y}_N)$, we can arrange
that $\# \mathcal{I} = P$ for any $P\in {1,2,\ldots,N}$, or in other words, that an observer near $(T,0)$ will see $P$ evolving density
fluctuations.
By part (i) and (ii), it then follows, by choosing $\ep \in (0,\ep_1]$ small enough, that not only does the observer see
the $P$  density fluctuations, but also that the evolution of gravitating fluid is well approximated on $M$ by
the cosmological Newtonian solution (\ref{cnsol.1}-\ref{cnsol.2}).

Taken together, the above results show that our solutions represent genuine cosmological solutions of the Einstein-Euler equations that are well-approximated on cosmological
scales by solutions of the cosmological Poisson-Euler equations. It is also clear from these results that these new solutions can be interpreted as a superposition of
local solutions, which have a local behavior similar to the solutions from \cite{Oliynyk:CMP_2010,Oliynyk:JHDE_2010}, that piece together into a consistent cosmological solution.
In this way, these new solutions overcome the deficiencies in the solutions from \cite{Oliynyk:CMP_2010,Oliynyk:JHDE_2010} that were identified in \cite{Green_Wald:2012}.

\sect{disc}{Discussion}

In this article, we describe the existence of a wide class of inhomogeneous relativistic solutions that
are well approximated by solutions to the cosmological Poisson-Euler equations of
Newtonian gravity on a cosmological spacetime region of the form $M=[0,T]\times \Rbb^3$. This rigourously
establishes that Newtonian gravity can serve as a good approximation to General Relativity on
cosmological scales. Moreover, the error estimates (\ref{crelest.1}-\ref{crelest.4}) provide the
precise sense in which the relativistic solutions
are approximated by Newtonian ones.

Although we have shown that it is possible to approximate relativistic cosmological solutions by Newtonian ones,
there remains much room for improvement. One key problem is to determine if the Newtonian solutions remain
a good approximation for large $T$, that is, in the limit  $T\rightarrow \infty$. It is conceivable that
there are secular effects that add up over time and cause the approximation to eventually break down. Another interesting
direction for investigation is to understand the most general class of initial data for which the approximation
continues to hold. For example, gravitational radiation could be present in the initial data as in \cite{Oliynyk:ATMP_2012} for the asymptotically flat case. Another direction would be to derive rigorous error estimates for the post-Newtonian corrections, i.e. higher order expansions in $\ep$, in order to capture relativistic effects. Yet another direction would be to include more sophisticated matter models
such as the Boltzmann equation to model particle scattering processes and also include electromagnetic effects
through coupling to the Maxwell field equations.

\bigskip

\begin{acknowledgments}
This work was partially supported by the ARC grants DP1094582 and FT1210045. The motivation to write
this article arose from discussions I had with R.W. Wald and S.R. Green
during a visit to the University of Chicago. I thank them for their
generosity with their time and many productive
and illuminating discussions. I also thank the referees for their comments
and criticisms, which have served to improve the content
and exposition of this article.
\end{acknowledgments}

\bibliography{cppnref}

\end{document}